\documentclass[twocolumn,showpacs,preprintnumbers,amsmath,amssymb,pra]{revtex4}
\usepackage{graphicx}
\usepackage{dcolumn}

\begin{document}
\title{Effect of phase fluctuation and dephasing on the dynamics of entanglement generation in a correlated emission laser}

\author{Sintayehu Tesfa}
\affiliation{Max Planck Institute for the Physics of Complex Systems, N$\ddot{o}$thnitzer Str. 38, 01187 Dresden, Germany\\Physics Department, Dilla University, P. O. Box 419, Dilla, Ethiopia}

\date{\today}

\begin{abstract} A detailed study of the effects of  phase fluctuation and dephasing on the dynamics of the entanglement generated from a coherently pumped correlated emission laser is presented. It is found that the time evolution of the entanglement is significantly reliant on the phase fluctuation and dephasing, particularly, at early stages of the lasing process. In the absence of external driving radiation, the degree of entanglement and intensity turns out to attain a maximum value just before starting to exhibit oscillation which dies at longer time scale. However, in case the driving mechanism is on, the oscillatory nature disappears due to the additional induced coherent superposition and the degree of entanglement would be larger at steady state. Moreover, the degree of entanglement as predicted by the logarithmic negativity and the Duan-Giedke-Cirac-Zoller criteria exhibits a similar nature when there is no driving radiation, although such a trend is eroded with increasing strength of the pumping radiation at longer time scale.  The other important aspect of the phase fluctuation and dephasing is the possibility of relaxing the time at which the maximum entanglement is detected.\end{abstract}

 \maketitle

 \section{INTRODUCTION}

In recent years, a nondegenerate three-level laser has attracted a great deal of interest in relation to its potential as a reliable source of robust entangled light \cite{pra74043816,jpb402373,prl94023601,pra75033816,jpb41145501,pra72022305}. The  nonclassical features of the radiation are predominantely attributed to the atomic coherence  induced by initially preparing the atoms in a coherent superposition of the energy levels between which a direct spontaneous transition is electric dipole forbidden \cite{pra74043816,jpb402373} or by pumping the same with an external radiation \cite{prl94023601,pra75033816,jpb41145501,pra72022305}. However, due to the fragility of the coherent superposition, the pertinent continuous variable entanglement is believed to be significantly ruined by various phenomena closely associated with the lasing process; noteworthy among many are dephasing and phase fluctuations. In practical situations, it is not possible to prepare the coherent superposition with an arbitrary perfection due to instability of the laser, faulty setups, and the inefficiency of the devices employed \cite{sinta,pra79013831,oc283781,pra444688,sint}. Even, in case one manages to prepare the coherent superposition somehow, it will subsequently decay due to the interaction with the surrounding \cite{pra79033810,pra79063815}. Therefore, a study of the effects of the phase fluctuations and dephasing on the evolution of the entanglement generation appears to be of a paramount importance.

Over the years, the issue of phase fluctuations has attracted a considerable attention \cite{p149}. In these studies the phase and photon number operators are presumed to fulfill a canonical commutation relation. In a more recent times, assuming the phase as randomly fluctuating $c$-number parameter, it is shown that the phase fluctuations substantially decrease the degree of entanglement \cite{pra77062308,pra79013831}. Investigation of a similar nature where the coherence is assumed to be partially prepared is also considered \cite{oc283781,sinta,sint}. In the present contribution, in order to investigate the effects of the phase fluctuations and dephasing on the time evolution of the generated entanglement in depth, coherently pumped correlated emission laser turns out to be a convenient tool (detailed description of the physical model can be found, for instance, in \cite{jpb41055503}). In this model, the three-level atoms are taken to be prepared initially in a partial 50:50 probability to be in the upper and lower energy levels. The atoms prepared in this manner are injected into the cavity at a constant rate and externally driven with a resonant coherent radiation while traversing the cavity. It is not difficult to envision that the pumping radiation could have also led to phase fluctuations; whose effect was thoroughly addressed elsewhere \cite{pra79013831,pra77062308,pra75062305}. Moreover, taking the recent advance in locking a phase in such a system into consideration, the rate at which the coherence superposition decays is set to be less than the spontaneous atomic decay rate. 

In the midest of the efforts geared towards  understanding   how to create, manipulate, explain and relate entanglement with various nonclassical correlations, it becomes evident that the degree of detectable entanglement when different criteria are used is found to be different. The emerging disparity can be fundamentally linked to the difference in the physical contexts and assumptions involved in deriving the correlations \cite{jpb41145504}. In view of this, over the years, quite a large volume of work aimed at quantifying the degree of available entanglement has appeared \cite{prl771413}. Essentially, the proposed criteria are in the form of some sort of inequality, in most instances provide only sufficient conditions and on top of that practical realization is mostly intractable due to the involved impairment in the corresponding measurement strategies. Therefore,   it would be imperative studying the nature of entanglement applying different criteria with the intention of looking for the overlap so that the likelihood of the prediction and realizeability becomes more reliable \cite{jpb42215506}. With this conviction, the time evolution of the entanglement is studied applying the criteria usually dubbed as logarithmic negativity \cite{pra65032314}, Duan-Giedke-Cirac-Zoller (DGCZ) \cite{prl842722} and Hillery-Zubairy (HZ) \cite{prl96050503}. The obtained results are compared whenever possible. 

In the study of the time evolution of a similar system when many parameters are involved, the usual approach is numerically calculating the required correlations making use of either the characteristic function \cite{pra79013831,pra72022305} or the rate equation \cite{pra77062308,oc283781}. However, in this article, the dynamics of the cavity radiation is analyzed following a straightforward analytic procedure recently outlined elsewhere \cite{pra74043816,pra79033810,sinta,sint}. The resulting analytical solutions were successfully employed earlier in the study of the time evolution of the two-mode squeezing and intensity \cite{sint}. Particularly, the time evolution of the two-mode squeezing shows that the phase fluctuation inherent with the imperfect preparation can enhance nonclassical features when the atoms are driven with a strong external radiation by creating spontaneous emission transition root via weakening the otherwise induced strong coupling of the upper and lower energy levels by pumping. However, in this contribution, it is shown that the degree of entanglement exhibits damping oscillatory nature when there is no driving radiation where in between there is  a narrow window of time in which the degree of entanglement is found to increase with the deviation of the phase fluctuation and rate of dephasing. The fluctuation of the degree of entanglement is solely attributed to the choice of the values of these parameters which are not completely independent.

\section{Equations of evolution}

The interaction of a pumped nondegenerate three-level cascade atom with a resonant two-mode cavity
radiation can be described in the rotating-wave approximation and
the interaction picture by the Hamiltonian of the form
\begin{align}\label{p01}\hat{H} &=ig[\hat{a}|a\rangle\langle
b| - |b\rangle\langle a|\hat{a}^{\dagger} +\hat{b}|b\rangle\langle
c| - |c\rangle\langle b|\hat{b}^{\dagger}]\notag\\& +
i{\Omega\over2}[|c\rangle\langle a|-|a\rangle\langle c|],\end{align}
where $\Omega$ is a real-positive constant proportional to the
amplitude of the driving radiation and $g$ is a coupling constant
chosen to be the same for both transitions. $\hat{a}$ and $\hat{b}$
are the annihilation operators that represent the two cavity modes. In the cascade configuration,
the transition from upper energy level $|a\rangle$ to the intermediate
level $|b\rangle$ and from level $|b\rangle$ to the lower energy level
$|c\rangle$ are taken to be resonant with the cavity radiation,
whereas the transition $|a\rangle\leftrightarrow|c\rangle$ is electric dipole forbidden.
Moreover, as a result of the emerging various quantum effects, it is assumed that the atoms can only be initially prepared in a partial maximum coherent superposition of the
upper and lower energy levels (a more general case has been presented in \cite{sinta}). 

On account of this, the initial state of the three-level atom can be written as
\begin{align}\label{p02}|\Psi_{A}(0)\rangle\; ={1\over\sqrt{2}}\big[|a\rangle +
e^{i\varphi}|c\rangle\big],\end{align} where  $\varphi$ is an arbitrary phase randomly distributed about a fixed mean value $\varphi_{0}$. Based on this assumption, the phase can be defined as $\varphi=\varphi_{0}+\delta\varphi$ in which $\delta\varphi$ is taken to be small random fluctuations around $\varphi_{0}$ that can be adjusted at will by proper choice of the phase of the cavity radiation \cite{oc283781}. In line with Eq. \eqref{p02}, the
corresponding initial reduced atomic density operator appears to be
\begin{align}\label{p03}\hat{\rho}_{A}(0)& =
{1\over2}\big[|a\rangle\langle a| + e^{-i\varphi}|a\rangle\langle
c| + e^{i\varphi}|c\rangle\langle a| +
|c\rangle\langle c|\big],\end{align} where
$e^{\pm i\varphi}/2$ stand for the initial atomic coherence. 

Addressing the contribution of every phase change does not seem practically realistic. Consequently, assuming the phase fluctuations as a Gaussian random process \cite{pra77062308,pra444688,oc283781} and using the deviation of the phase fluctuation instead of the actual phase led to more tractable situation \cite{sint}. With similar conviction, employing the property of the Gaussian variables \cite{method}, one can express
\begin{align}\label{p21}\langle\exp\pm i\delta\varphi\rangle=\exp-\langle\delta\varphi^{2}/2\rangle.\end{align} Moreover, for a Gaussian random process $\langle\delta\phi\rangle$ is zero where  $\langle\delta^{2}\phi/2\rangle=\theta$ represents the deviation which is generally designated as phase fluctuation. 

Furthermore, following the detailed proceduce of obtaining the master equation and equations of evolution presented elsewhere \cite{sinta,sint}, it is possible to verify that 
\begin{align}\label{p04}\alpha(t) = C_{+}(t)\alpha(0) + D_{+}(t)\beta^{*}(0) + E_{+}(t)
 + F_{+}(t),\end{align}
\begin{align}\label{p05}\beta(t) =  C_{-}(t)\beta(0)+D_{-}(t)\alpha^{*}(0)
+ E_{-}(t) + F_{-}(t),\end{align}
 where
\begin{align}\label{p06}C_{\pm}(t) = \frac{1}{2}\left[(1\pm p)e^{-\mu_{-}t} +
(1\mp p)e^{-\mu_{+}t}\right],\end{align}
\begin{align}\label{p07}D_{\pm}(t) = \frac{q_{\pm}}{2}[e^{-\mu_{+}t} - e^{-\mu_{-}t}],\end{align}
\begin{align}\label{p08}E_{+}(t)& =  \frac{1}{2}\int_{0}^{t}[(1+p)e^{-\mu_{-}(t-t')} \notag\\&+
(1-p)e^{-\mu_{+}(t-t')}]f_{a}(t')dt',\end{align}
\begin{align}\label{p09}E_{-}(t) &= \frac{1}{2}\int_{0}^{t}[(1-p)e^{-\mu_{-}(t-t')}
\notag\\&+ (1+p)e^{-\mu_{+}(t-t')}]f_{b}(t')dt',\end{align}
\begin{align}\label{p10}F_{+}(t) = \frac{q_{+}}{2}\int_{0}^{t}[e^{-\mu_{+}(t-t')} -
e^{-\mu_{-}(t-t')}]f^{*}_{b}(t')dt',\end{align}
\begin{align}\label{p11}F_{-}(t) = \frac{q_{-}}{2}\int_{0}^{t}[e^{-\mu_{+}(t-t')} -
e^{-\mu_{-}(t-t')}]f_{a}^{*}(t')dt',\end{align} with
\begin{align}\label{p12}\mu_{\pm} &= {\kappa\over2}+{A\over2B}\left[(2\zeta'+\zeta)e^{-\theta}\right.\notag\\&\left.\pm\sqrt{\zeta'^{2}(1+\zeta\zeta')^{2}+4(\zeta'^{2}+\chi)^{2}-
[(2-\zeta'\zeta)e^{-\theta}]^{2}}\right],\end{align}
\begin{align}\label{p13}p={2\big[\zeta'^{2}+\chi\big]\over
\sqrt{\zeta'^{2}(1+\zeta\zeta')^{2}+4(\zeta'^{2}+\chi)^{2}-
[(2-\zeta'\zeta)e^{-\theta}]^{2}}},\end{align}
\begin{align}\label{p14}q_{\pm}={-\zeta'(1+\zeta'\zeta)\pm\big[(2-\zeta'\zeta)e^{-\theta}\big]
\over
\sqrt{\zeta'^{2}(1+\zeta\zeta')^{2}+4(\zeta'^{2}+\chi)^{2}-
[(2-\zeta'\zeta)e^{-\theta}]^{2}}},\end{align} in which
\begin{align}\label{p15}A = \frac{2r_{a}g^{2}}{\gamma^{2}},\end{align}
\begin{align}\label{p16}B=(4+\zeta^{2})(1+\zeta'\zeta),\end{align} $\zeta={\Omega/\gamma}$, $\zeta'={\Omega/\Gamma}$, and $\chi=\gamma/\Gamma.$ It is worth noting that $\kappa$ is the cavity damping constant, $\Gamma$ is the spontaneous atomic damping rate, $\gamma$ is the rate at which the coherent superposition decays and $r_{a}$ is the rate at which the atoms are injected into the cavity. 

In addition, the corresponding stochastic noise forces are shown to have  correlations of the form
\begin{align}\label{p17}\langle f_{a}(t')f^{*}_{a}(t)\rangle & = {A\over B}\left[2\zeta'^{2}+2\chi-(2\zeta'+\zeta)e^{-\theta}\right] \delta(t-t'),\end{align}
\begin{align}\label{p18}\langle f_{b}(t')f^{*}_{b}(t)\rangle = 0,\end{align}
 \begin{align}\label{p19}\langle f_{b}(t')f_{a}(t)\rangle&=
 \frac{A}{2B}\left[\zeta'(1+\zeta'\zeta)+(2-\zeta'\zeta)e^{-\theta}\right]\delta(t-t'),\end{align}
 \begin{align}\label{p20}\langle f^{*}_{b}(t')f_{a}(t)\rangle=\langle f_{a}(t')f_{a}(t)\rangle
 =\langle f_{b}(t')f_{b}(t)\rangle = 0.\end{align}

\section{Continuous variable entanglement}

In recent years, on the basis of different conditions and assumptions, several inseparability criteria for continuous variable product states have been proposed \cite{prl842722,prl96050503,pla2231}. One of these is related to a logarithmic negativity ($E_{N}$) \cite{job7577,pra65032314,pra70022318} which can be defined as \cite{pra70022318,prl93063601,jpb41215502}
\begin{align}\label{p22}E_{N}=\rm{max}[0,-\log_{2}V_{s}],\end{align} where $V_{s}$ is the smallest eigenvalue of the symplectic matrix \cite{pra70022318,prl93063601}. The inseparability condition associated with this approach is based on the well established fact that for the composite state to be separable to its constitute, the product density operator should have a positive partial transpose.

In light of this, upon solving the eigenvalue equations for the symplectic spectrum of the covariance matrix of the partially transposed density operator, the smallest eigenvalue is found to have a form 
\begin{align}\label{p23}V_{s}=\left[\frac{\xi-\sqrt{\xi^{2}-\rm4det\Xi}}{2}\right]^{1/2},\end{align} with
\begin{align}\label{p24}\xi= \rm{det}\sigma_{A}+det\sigma_{B}-2det\sigma_{AB},\end{align}
where $\sigma_{A}$ and $\sigma_{B}$ are the covariance matrices describing each modes separately while $\sigma_{AB}$ represents the intermodal correlations \cite{pra65032314}.

In connection with the bimodal states involved, the $2X2$ block form of the covariance matrix $\Xi$ can be expressed as \cite{pra79032334}
\begin{align}\label{p25}\Xi=\left(
                         \begin{array}{cc}
                           \sigma_{A} & \sigma_{AB} \\
                           \sigma_{AB}^{T} & \sigma_{B} \\
                         \end{array}
                       \right),
\end{align} where
\begin{align}\label{p26}\Xi_{ij}={1\over2}\langle\hat{X}_{i}\hat{X}_{j}+\hat{X}_{j}\hat{X}_{i}\rangle -\langle\hat{X}_{i}\rangle\langle\hat{X}_{j}\rangle.\end{align} The corresponding quadrature operators are defined as
$\hat{X}_{1}=\hat{a}+\hat{a}^{\dagger}$, $\hat{X}_{2}=i(\hat{a}^{\dagger}-\hat{a})$,
$\hat{X}_{3}=\hat{b}+\hat{b}^{\dagger}$ and
$\hat{X}_{4}=i(\hat{b}^{\dagger}-\hat{b})$. With this introduction, the covariance matrix can be expanded as
\begin{align}\label{p27}\Xi=\left(
                         \begin{array}{cccc}
                           m & 0 & c & 0 \\
                           0 & m & 0 & -c \\
                           c & 0 & n & 0 \\
                           0 & -c & 0 & n \\
                         \end{array}
                       \right),
\end{align} where 
\begin{align}m&=1+2\langle\hat{a}^{\dagger}(t)\hat{a}(t)\rangle+\langle\hat{a}^{2}(t)\rangle+\langle\hat{a}^{\dagger^{2}}(t)\rangle\notag\\&
-[\langle\hat{a}(t)\rangle^{2}+\langle\hat{a}^{\dagger}(t)\rangle^{2}+2\langle\hat{a}(t)\rangle\langle\hat{a}^{\dagger}(t)\rangle],\end{align}
\begin{align}n&=1+2\langle\hat{b}^{\dagger}(t)\hat{b}(t)\rangle+\langle\hat{b}^{2}(t)\rangle+\langle\hat{b}^{\dagger^{2}}(t)\rangle\notag\\&
-[\langle\hat{b}(t)\rangle^{2}+\langle\hat{b}^{\dagger}(t)\rangle^{2}+2\langle\hat{b}(t)\rangle\langle\hat{b}^{\dagger}(t)\rangle],\end{align}
\begin{align}c&=\langle\hat{a}(t)\hat{b}(t)\rangle +\langle\hat{a}^{\dagger}(t)\hat{b}^{\dagger}(t)\rangle+\langle\hat{a}(t)\hat{b}^{\dagger}(t)\rangle +\langle\hat{a}^{\dagger}(t)\hat{b}(t)\rangle
\notag\\&-[\langle\hat{a}(t)\rangle\langle\hat{b}(t)\rangle +\langle\hat{a}^{\dagger}(t)\rangle\langle\hat{b}^{\dagger}(t)\rangle+\langle\hat{a}(t)\rangle\langle\hat{b}^{\dagger}(t)\rangle \notag\\&+\langle\hat{a}^{\dagger}(t)\rangle\langle\hat{b}(t)\rangle].\end{align} These equations can be expressed in terms of $c$-number variables associated with the normal ordering as 
\begin{align}\label{p28}m=1+2\langle\alpha^{*}(t)\alpha(t)\rangle,\end{align}
\begin{align}\label{p29}n=1+2\langle\beta^{*}(t)\beta(t)\rangle,\end{align}
\begin{align}\label{p30}c=\langle\alpha(t)\beta(t)\rangle +\langle\alpha^{*}(t)\beta^{*}(t)\rangle.\end{align} In obtaining these expressions, upon neglecting the interaction among the atoms in the cavity and assuming the cavity to be initially in a vacuum state, it is taken that \cite{sint}
$$\langle\alpha(t)\rangle=\langle\beta(t)\rangle=\langle\alpha^{2}(t)\rangle=\langle\beta^{2}(t)\rangle=\langle\alpha(t)\beta^{*}(t)\rangle=0.$$ 

Furthermore, making use  of Eqs. \eqref{p25}, \eqref{p26}, \eqref{p27}, \eqref{p28}, \eqref{p29} and \eqref{p30}, one can readily see that
\begin{align}\label{p31}\rm det\sigma_{A}&=1+4\langle\alpha^{*}(t)\alpha(t)\rangle[\langle\alpha^{*}(t)\alpha(t)\rangle+1],\end{align}
\begin{align}\label{p32}\rm det\sigma_{B}&=1+4\langle\beta^{*}(t)\beta(t)\rangle[\langle\beta^{*}(t)\beta(t)\rangle+1],\end{align}
\begin{align}\label{p33}\rm det\sigma_{AB}&=-4\langle\alpha(t)\beta(t)\rangle^{2},\end{align}
\begin{align}\label{p34}\rm det \Xi &= 16 \left[{1\over4}+{\langle\alpha^{*}(t)\alpha(t)\rangle+\langle\beta^{*}(t)\beta(t)\rangle\over2}\right.\notag\\&\left.
+\langle\alpha^{*}(t)\alpha(t)\rangle\langle\beta^{*}(t)\beta(t)\rangle-\langle\alpha(t)\beta(t)\rangle^{2}\right]^{2}.\end{align}

It is a well established fact that the two-mode Gaussian product state would be entangled if $\rm E_{N}=-Log_{2}V_{s}$ which implies that $\rm Log_{2}V_{s}$ should be negative. That means
\begin{align}\label{p35}V_{s}<1\end{align} represents the condition for detecting entanglement. In order to quantify entanglement using this approach directly, the  various correlations in Eq. \eqref{p34} should be
determined applying Eqs. \eqref{p04} and \eqref{p05}. To this end, assuming the
cavity to be initially in a two-mode vacuum state and the noise force at time $t$ does not
statistically related to the cavity mode variables at earlier
times, one can
readily verify that \cite{sint}
\begin{align}\label{p37}\langle\alpha^{*}(t)\alpha(t)\rangle &=
A\left[\frac{L(1-p)^{2}+
Mq_{+}(1-p)}{8B\mu_{+}}\right][1 -
e^{-2\mu_{+}t}] \notag\\&+A\left[\frac{L(1+p)^{2}- Mq_{+}(1+p)}{8B\mu_{-}}\right][1-e^{-2\mu_{-}t}]
\notag\\&+A\left[ \frac{L(1-p^{2}) +Mq_{+}p}{2B(\mu_{+}+\mu_{-})}\right][1
- e^{-(\mu_{+}+\mu_{-})t}],\end{align}
\begin{align}\label{p38}\langle\beta^{*}(t)\beta(t)\rangle &= A\left[\frac{Lq_{-}^{2}
+Mq_{-}(1+p)}{8B\mu_{+}}\right][1 - e^{-2\mu_{+}t}]
\notag\\&+A\left[\frac{Lq_{-}^{2} -Mq_{-}(1-p)}{8B\mu_{-}}\right][1-e^{-2\mu_{-}t}]
\notag\\&-A\left[ \frac{Lq_{-}^{2}+ Mq_{-}p}{2B(\mu_{+}+\mu_{-})}\right][1 -
e^{-(\mu_{+}+\mu_{-})t}],\end{align}
\begin{align}\label{p39}\langle\alpha(t)\beta(t)\rangle &= A\left[\frac{2Lq_{-}(1-p)+M(1-p^{2}+q_{-}q_{+})}{16B\mu_{+}}\right]\notag\\&\times[1
- e^{-2\mu_{+}t}] \notag\\&-A\left[\frac{2Lq_{-}(1+p)- M(1-p^{2}+q_{-}q_{+})}{16B\mu_{-}}\right]\notag\\&\times[1-e^{-2\mu_{-}t}]
\notag\\&+A\left[ \frac{2Lq_{-}p +M(1+p^{2}-q_{-}q_{+})}{4B(\mu_{+}+\mu_{-})}\right]\notag\\&\times[1
- e^{-(\mu_{+}+\mu_{-})t}],\end{align}
where
\begin{align}\label{p41}L=2\zeta'^{2}+2\chi-(2\zeta'+\zeta)e^{-\theta},\end{align}
\begin{align}\label{p42}M=\zeta'(1+\zeta'\zeta)+(2-\zeta'\zeta)e^{-\theta}.\end{align}

In the following, the time evolution of the degree of detectable entanglement is investigated employing the criterion of logarithmic negativity \eqref{p35}. To achieve the intended goal, $V_{s}$ is plotted against $time$ (a scaled unitless parameter) applying Eqs. \eqref{p23}, \eqref{p24}, \eqref{p31}, \eqref{p32}, \eqref{p33}, \eqref{p34}, \eqref{p37}, \eqref{p38}, \eqref{p39}, \eqref{p41} and \eqref{p42}. In order to see the nature of entanglement in depth, various special cases are considered.

\subsection{When there is no external driving radiation}

In recent years, the investigation of the quantum features of the radiation generated by the nondegenerate three-level cascade laser when $\gamma=\Gamma$ and $\theta=0$ has attracted interest. These studies are usually confined to the steady state case \cite{pra74043816,pra77013815} or numerical solution  for selected parameters \cite{pra72022305,pra77062308}. However, in this article, upon using the rate of dephasing and deviation of the phase fluctuation as a probe, the evolution of the entanglement is studied when $\Omega=0$.

\begin{figure}[htb]
 \centerline{\includegraphics [height=6.5cm,angle=0]{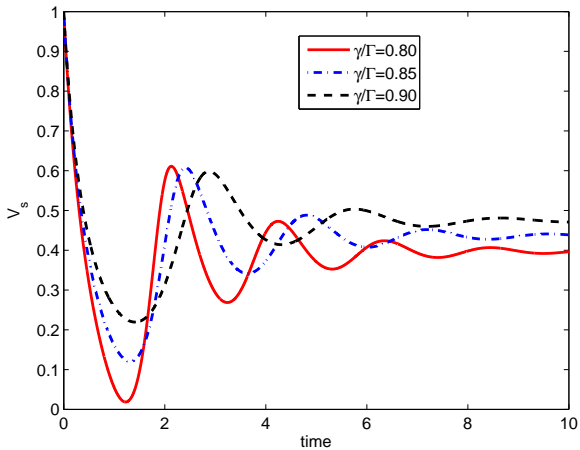}}
 \caption {\label{fig1} Plots of the time evolution of the smallest eigenvalue ($V_{s}$) of the cavity radiation for $\kappa = 0.5$, $\theta=0$, $A=10$, $\Omega=0$ and different values of $\gamma/\Gamma$.}
 \end{figure}

 \begin{figure}[htb]
 \centerline{\includegraphics [height=6.5cm,angle=0]{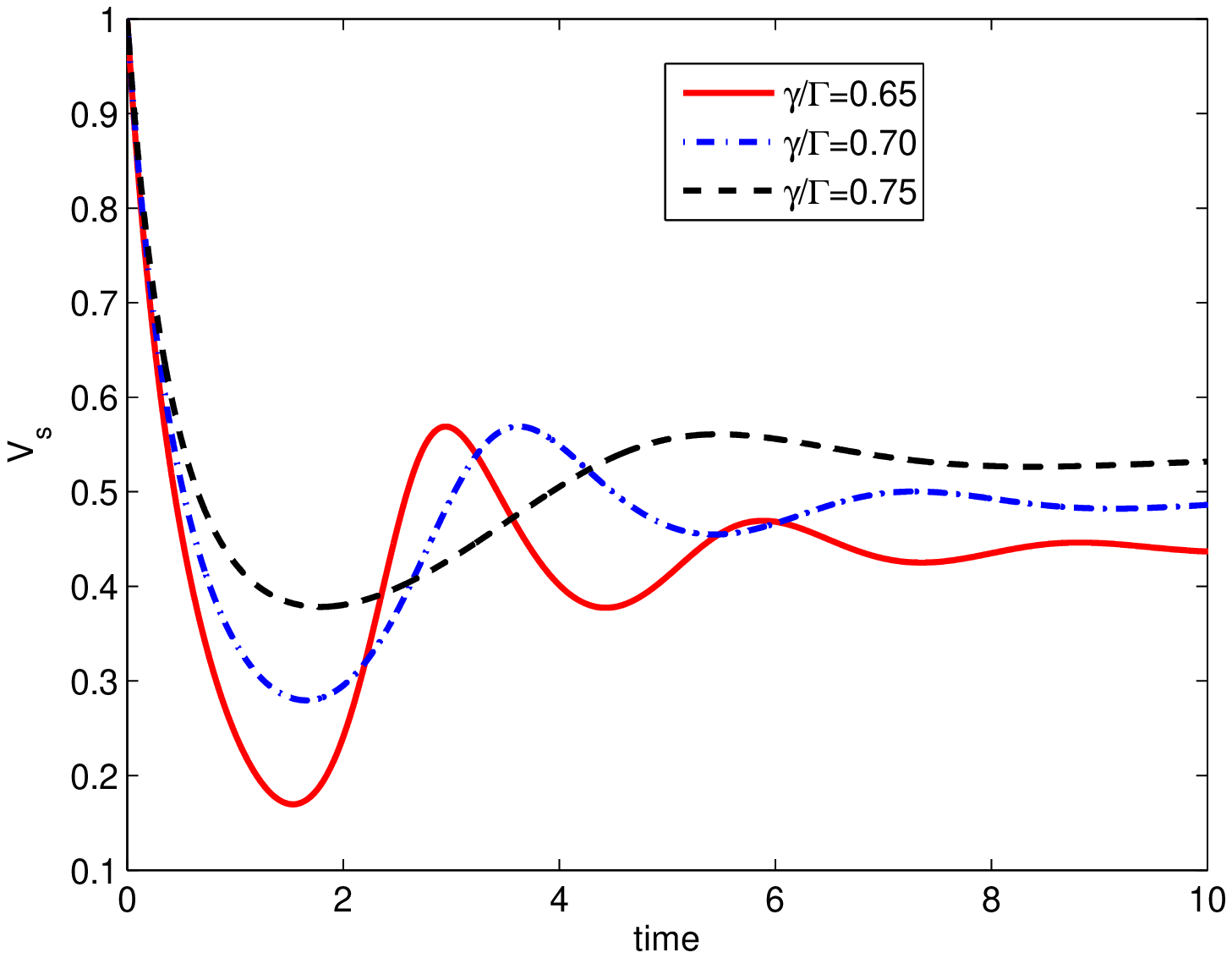}}
 \caption {\label{fig2} Plots of the time evolution of the smallest eigenvalue ($V_{s}$) of the cavity radiation for $\kappa = 0.5$, $\theta=0.25$, $A=10$, $\Omega=0$ and different values of $\gamma/\Gamma$.}
 \end{figure}

\begin{figure}[htb]
 \centerline{\includegraphics [height=6.5cm,angle=0]{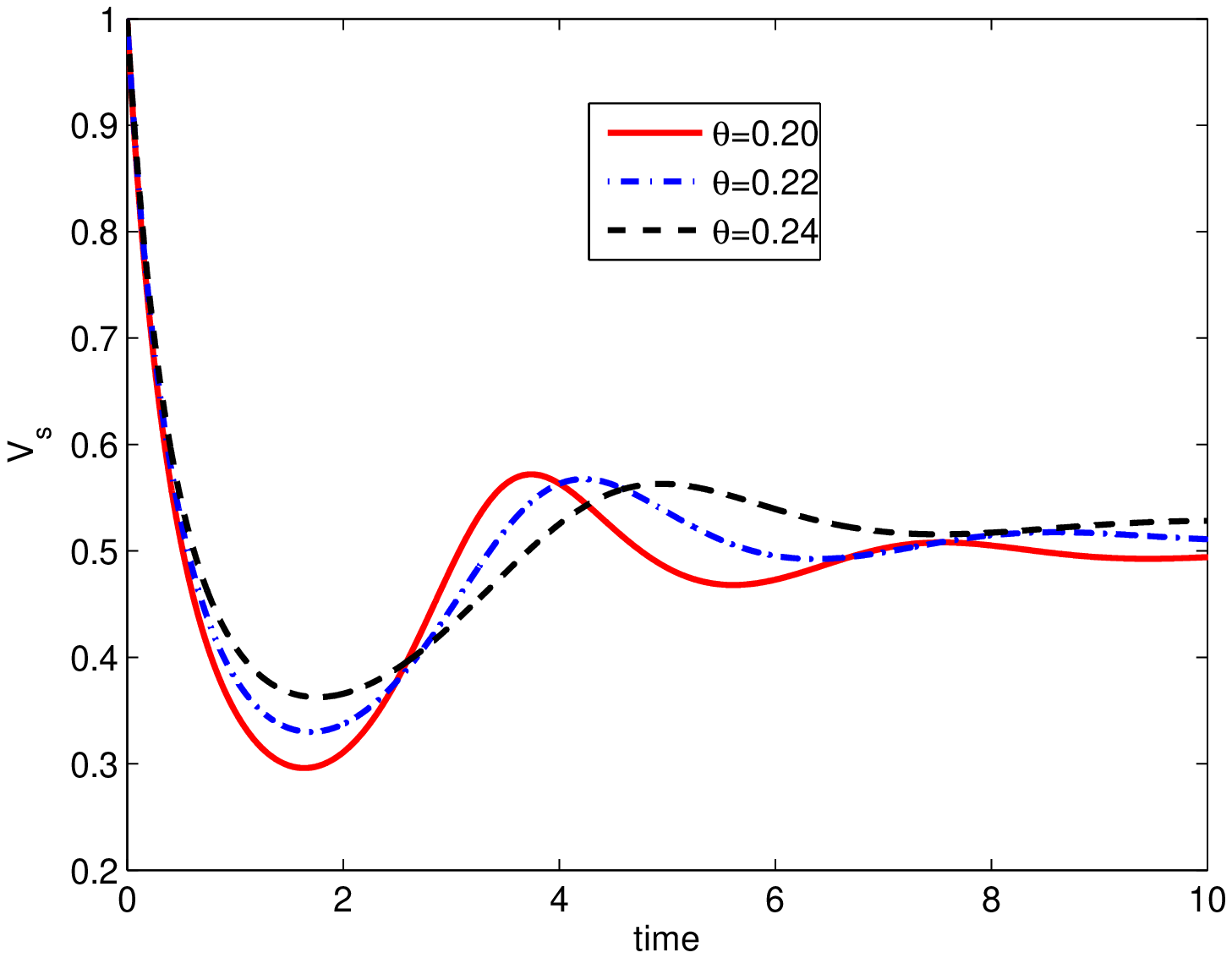}}
 \caption {\label{fig3} Plots of the time evolution of the smallest eigenvalue ($V_{s}$) of the cavity radiation for $\kappa = 0.5$, $\gamma=0.75\Gamma$, $A=10$, $\Omega=0$ and different values of $\theta$.}
 \end{figure}

\begin{figure}[htb]
 \centerline{\includegraphics [height=6.5cm,angle=0]{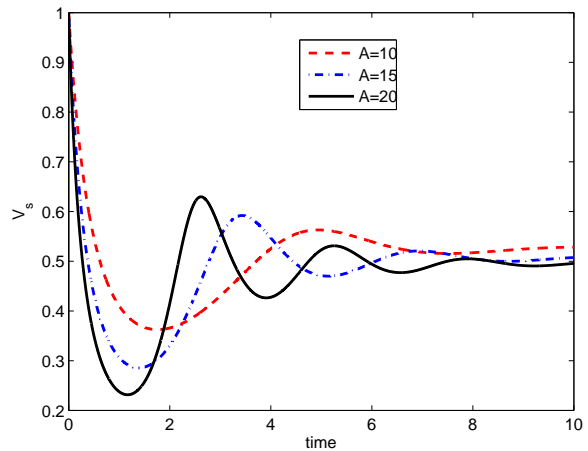}}
 \caption {\label{fig4} Plots of the time evolution of the smallest eigenvalue ($V_{s}$) of the cavity radiation for $\kappa = 0.5$, $\theta=0.25$, $\gamma=0.75\Gamma$, $\Omega=0$ and different values of $A$.}
 \end{figure}

As clearly shown in Figs. \ref{fig1}, \ref{fig2}, \ref{fig3} and \ref{fig4},  the degree of entanglement exhibits damping oscillatory nature for certain parameters under consideration. It is worth noting that, since the injection of the atoms is launched at $t=0$, entanglement would not be expected at the beginning when the cavity is taken to be initially at vacuum state. Moreover, it is not difficult to see that the degree of entanglement increases rapidly and then starts to fluctuate with time. The increment of the degree of entanglement at early stages of the lasing process can be directly linked to the increment of the atoms that participate in the correlated emission. Similar increment behavior at early stages of lasing operation was observed for various schemes of three-level laser under different conditions using various entanglement criteria \cite{pra72022305,pra79013831,pra444688,oc283781}. 

Nevertheless, in order to trace the origin of the fluctuation, it appears natural referring back to Eq. \eqref{p12}. In Eq. \eqref{p12}, if $\Omega=0$, the term under square root reduces to $\sqrt{4(\chi^{2}-e^{-2\theta})}$. This indicates that, in Eqs. \eqref{p37}, \eqref{p38} and \eqref{p39}, the exponential term that contains time would be complex valued which leads to oscillation if $\gamma/\Gamma<e^{-\theta}$. In case of $\theta=0$ and $\gamma=\Gamma$, generally the oscillation in the entanglement evolution is not observed as large volume of earlier works asserted \cite{pra72022305,pra79013831,pra444688,oc283781,pra77062308,sint}. One can hence deduce that the main source of this oscillatory nature is the competition between the rate at which the coherent superposition decays and the degree at which the same coherence exhibits fluctuations initially. Quite obviously, this competition can be directly associated with the mean photon number of the generated radiation. Although it is not presented in this work, the pertinent mean photon number also exhibits a similar kind of oscillatory nature. Therefore, comparison between the two reveals that the degree of entanglement and the mean photon number are larger for the same parameters. It is hence anticipated that the thermal fluctuations arising due to high intensity might be responsible for the downturn of the quantum features (increment of $V_{s}$) \cite{sint}. It is worth noting at this juncture that in the definition of the covariance matrix,  terms related to normal ordering are included that are basically associated with the contribution of vacuum fluctuations which might have played a significant role. 

It is not hard to notice from Fig. \ref{fig1} that the smallest eigenvalue corresponding to the  covariance matrix is smaller for smaller values of $\gamma/\Gamma$ at early stages of the lasing process. This indicates that the degree of entanglement by large decreases with the increasing rate at which the coherence superposition decays provided that the rate at which the atoms spontaneously decay is taken to be constant. However, it would be appropriate noting that for small pockets of time, the degree of entanglement can increase with the rate at which the coherence decays. For relatively longer time span, the degree of entanglement slowly oscillates very close to $V_{s}=0.5$. This can be fairly related to the earlier report for the same system when $\gamma=\Gamma$, $\eta=0$, $\theta=0$ and $A=10$, where $V_{s}$ is found to be 0.5 at steady state \cite{jpb42215506}. On the other hand, critical scrutiny of Figs \ref{fig1} and \ref{fig2} reveals that the time at which the degree of entanglement becomes maximum increases with the rate of dephasing. Though increasing the rate of dephasing significantly damages the generated entanglement, it can be utilized in prolonging the time at which the maximum entanglement can be generated. This can be taken as a positive aspect in practical realization of entanglement in such a system where the corresponding actual time is very small. 

In plotting Fig. \ref{fig2}, two changes are made; the values of $\gamma/\Gamma$ are reduced while the value of $\theta$ is increased so that the nature of the evolution of entanglement be clearly evident. If one expects the nature of entanglement to follow the same trend for various values of $\gamma/\Gamma$, the degree of entanglement indicated in Fig. \ref{fig2} should have been much larger. However, when the results presented in Figs. \ref{fig1} and \ref{fig2} are compared, the entanglement turns out to be better in the former. This entails that the phase fluctuation significantly ruins the obtainable degree of entanglement at earlier times. For relatively longer time span, the oscillation is limited to very close to $V_{s}=0.5$ as in the case when $\theta=0$. This suggests that the effect of phase fluctuation can be minimal at longer time scale when compared to earlier times. In order to attest to this claim, $V_{s}$ is plotted against $time$ for different $\theta$. The result depicted in Fig \ref{fig3} indicates that, indeed, the degree of entanglement decreases with the deviation of the phase fluctuation in the earlier times of the lasing operation. In relation to the involved oscillation, there is also a  brief window of time in which the degree of entanglement can increase with the phase fluctuation. Just as we have observed for the rate of dephasing, the time at which the maximum entanglement can be detected increases with the phase fluctuation.

Aiming at the prospect of increasing the degree of entanglement, $V_{s}$ is plotted against $time$ for $\theta=0.25$, $\gamma=0.75\Gamma$ and different values of the linear gain coefficient. As clearly shown in Fig. \ref{fig4}, the degree of entanglement increases with the linear gain coefficient in the early stages of the lasing process. Quite remarkably, in light of the oscillation, there is a window of time where the entanglement can decrease with the linear gain coefficient. Once again, at larger time scale the dependence of the entanglement on the linear gain coefficient tends to be much lesser when compared to the earlier times. Fitting things together may indicate that the nonclassical features and intensity of the cavity radiation take maximum values before they start damping oscillation that dies at longer time scale where the steady state properties begins to take effect.

\subsection{In stronger driving regime}

The entanglement generated from coherently pumped correlated emission laser when $\eta=0$, $\gamma=\Gamma$ and $\theta=0$ was studied earlier using the criteria of DGCZ. It is generally noted  that $\Delta u^{2}+\Delta v^{2}$ is very close to one at steady state in a strong driving limit ($\Omega>>\Gamma$) \cite{jpb41055503}. In the following, upon varying the values of the rate of dephasing ($\gamma/\Gamma$) and the deviation of the phase fluctuation ($\theta$), the time evolution of the entanglement that can be detected by the logarithmic negativity is studied.

\begin{figure}[htb]
 \centerline{\includegraphics [height=6.5cm,angle=0]{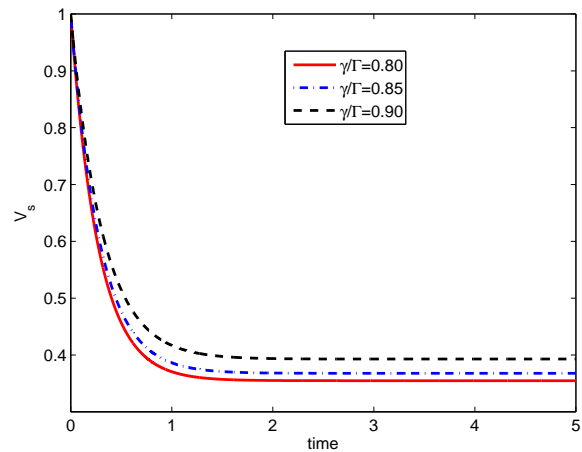}}
 \caption {\label{fig5} Plots of the time evolution of the smallest eigenvalue ($V_{s}$) of the cavity radiation for $\kappa = 0.5$, $\theta=0$, $A=10$, $\Omega=10\Gamma$ and different values of $\gamma/\Gamma$.}
 \end{figure}

\begin{figure}[htb]
 \centerline{\includegraphics [height=6.5cm,angle=0]{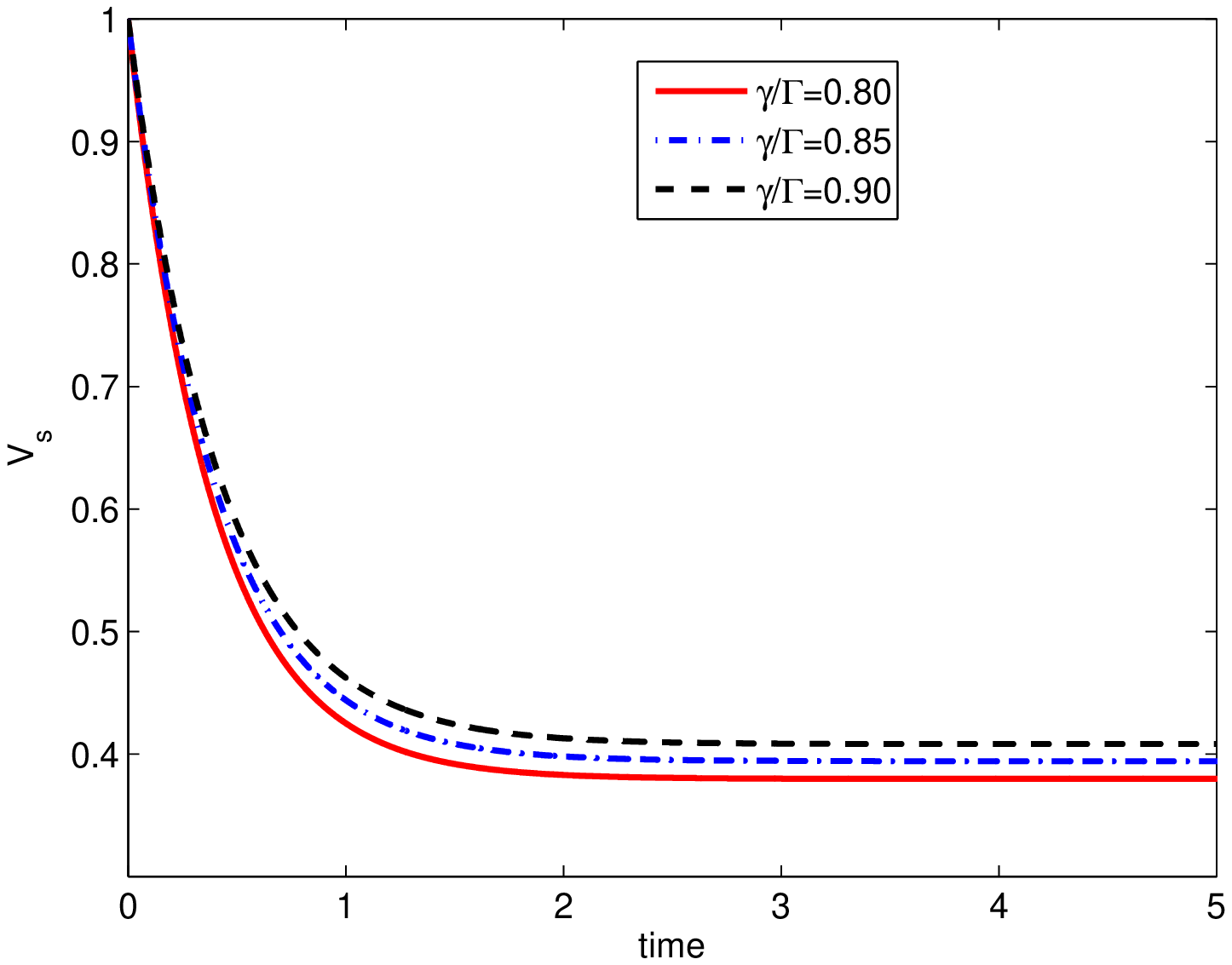}}
 \caption {\label{fig6} Plots of the time evolution of the smallest eigenvalue ($V_{s}$) of the cavity radiation for $\kappa = 0.5$, $\theta=0.25$, $A=10$, $\Omega=10\Gamma$ and different values of $\gamma/\Gamma$.}
 \end{figure}

\begin{figure}[htb]
 \centerline{\includegraphics [height=6.5cm,angle=0]{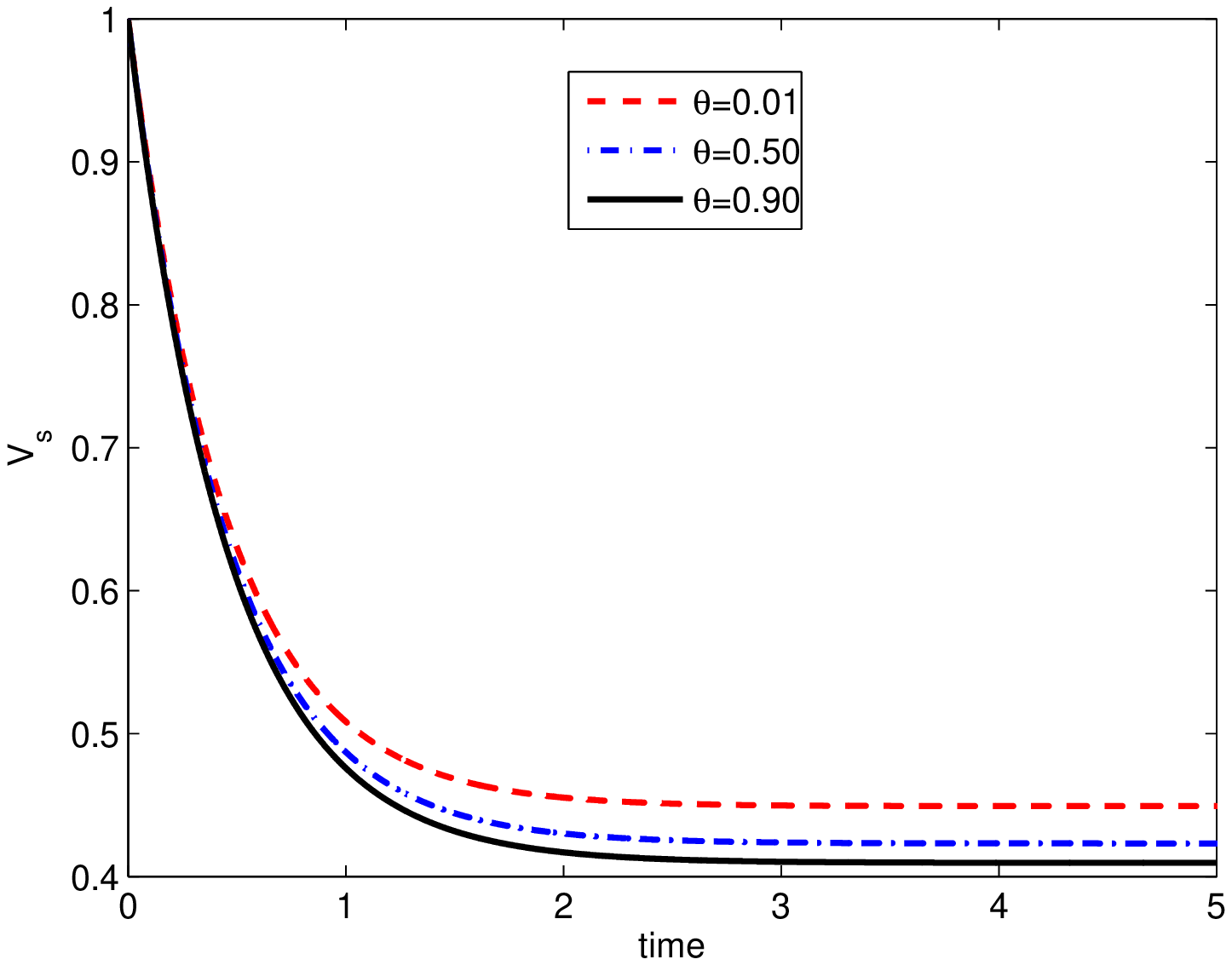}}
 \caption {\label{fig7} Plots of the time evolution of the smallest eigenvalue ($V_{s}$) of the cavity radiation for $\kappa = 0.5$, $\gamma=\Gamma$, $A=10$, $\Omega=10\Gamma$ and different values of $\theta$.}
 \end{figure}

\begin{figure}[htb]
 \centerline{\includegraphics [height=6.5cm,angle=0]{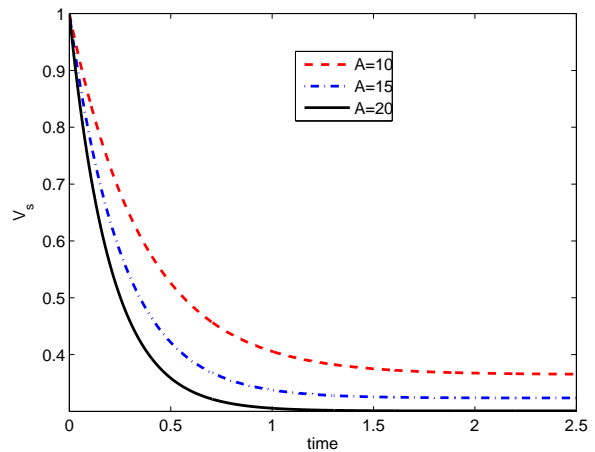}}
 \caption {\label{fig8} Plots of the time evolution of the smallest eigenvalue ($V_{s}$) of the cavity radiation for $\kappa = 0.5$, $\gamma=0.75\Gamma$, $\theta=0.25$, $\Omega=10\Gamma$ and different values of $A$.}
 \end{figure}

In line with earlier discussion, the term under square root in Eq. \eqref{p12} would not be negative for large $\Omega$. As a result, as clearly evinced in Figs. \ref{fig5}, \ref{fig6}, \ref{fig7} and \ref{fig8}, the degree of entanglement does not exhibit oscillatory nature in the strong driving limit. It is, rather, observed that the degree of entanglement increases with time for all parameters under consideration. This outcome suggests that a better degree of entanglement would be expected at steady state which is believed to be an important aspect in practical utilization of the system. It can readily be seen from Fig. \ref{fig5} that when the atoms are pumped externally with  strong coherent radiation, the entanglement appears to decrease with the rate of dephasing. Moreover, one can observe from Fig. \ref{fig6} that the nature of the degree of entanglement does not display much change from what is imparted in Fig. \ref{fig5} except decrement in the effect of the rate of phase fluctuation slightly. In order to show the dependence of the evolution of the entanglement on the phase fluctuation in depth, $V_{s}$ is plotted against $time$ for different  values of $\theta$. Relatively spaced values of $\theta$ are chosen so that the modification in the evolution of the entanglement  be evident from the plots. As can easily be noticeable from Fig. \ref{fig7}, the degree of entanglement increases with the phase fluctuation. This is in a complete agreement with a recent observation that in case when there is a strong external driving the phase fluctuation enhances nonclassical features via creation of indirect spontaneous emission root \cite{sint}. 

It is not difficult to see from Figs. \ref{fig5}, \ref{fig6} and \ref{fig7} that $V_{s}$ is very close to 0.4, which is by far larger than the case when $\Omega=0$. In the quest of improving the degree of entanglement, $V_{s}$ is plotted against $time$ for $\gamma=0.75\Gamma$, $\theta=0.25$ and different values of $A$. As unambiguously presented in Fig. \ref{fig8}, the degree of entanglement is found to increase with the linear gain coefficient. This implies that in the process of practical utilization, one can manipulate the rate at which the atoms are injected into the cavity to get an optimum entanglement. It is worth noting that the values of $A$ are set in this manner so that one can able to make comparison with the case when $\Omega=0$ rather than aiming at finding the highest degree of entanglement. In light of this, upon comparing the results shown in Figs. \ref{fig4} and \ref{fig8}, one can readily see that the degree of entanglement is larger when $\Omega=0$ at earlier times, whereas the opposite holds for longer time span where the oscillation naturally subsides. Although there is no earlier analysis that supports this outcome, the result obtained here entails that the driving mechanism is required if one wishes to generate strongly entangled light at steady state for the system under consideration.

\section{Comparison of the degree of entanglement}

It is a well known fact that the criteria used to detect the degree of entanglement of the continuous variable composite system are of different kind and also related to different physical contexts.  It so happens that generally they do not lead to similar conclusion since predominantly they are emanated from different principles and assumptions. In order to get a clear understanding of the nature of entanglement, it would be necessary studying the relation among the predictions of various criteria. With this understanding, in the following, the evolution of the entanglement that was obtained applying the criterion of logarithmic negativity is compared with the DGCZ and HZ criteria.

\subsection{Duan-Giedke-Cirac-Zoller criteria}

In this section, the nature of entanglement as predictable by the criterion following from the variance of the Einstein-Podolsky-Rosen (EPR) type quadrature operators \cite{pr47777} is compared with logarithmic negativity that is employed in the preceding section of this work. Previous study shows that the nature of the predicted degree of entanglement by the two criteria for a similar system when $\Omega=0$, $\theta=0$ and $\gamma=\Gamma$ has the same form except near $\eta=0$ where significant disparity was observed  at steady state \cite{jpb42215506}. The main aim here is shifted to the study of the time evolution of the entanglement upon using $\gamma/\Gamma$ and $\theta$ as a probe. It is worth noting that the detailed analysis of the degree of entanglement that can be  quantified by the criterion following from DGCZ has been presented in connection with the degree of two-mode squeezing elsewhere \cite{sint}. In earlier work, the oscillatory nature of the entanglement was not observed since the amplitude of the driving radiation was set to be $\Omega\ge0.5\Gamma$. With the aid of the calculation presented earlier, the variance of the quadrature operators ($\Delta u^{2}+\Delta v^{2}$) and the smallest eigenvalue of the symplectic matrix ($V_{s}$) are plotted against $time$ for various parameters. In the plot ($\Delta u^{2}+\Delta v^{2}$) is multiplied by half so that comparison can be as straightforward as possible.

\begin{figure}[htb]
 \centerline{\includegraphics [height=6.5cm,angle=0]{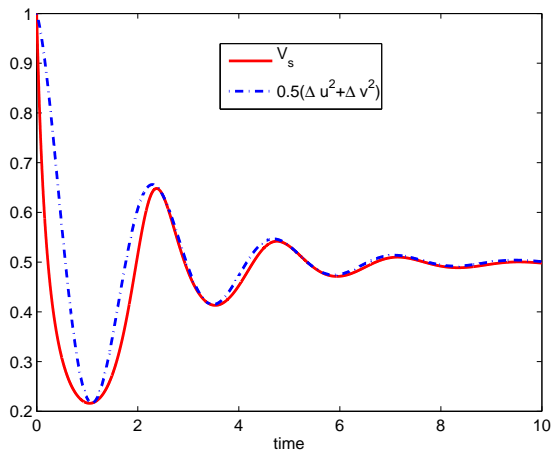}}
 \caption {\label{fig9} Plots of the time evolution of the smallest eigenvalue ($V_{s}$) and variance of the EPR operators ($\Delta u^{2}+\Delta v^{2}$) of the cavity radiation for $\kappa = 0.5$, $\theta=0.25$, $A=25$, $\Omega=0$ and $\gamma=0.75\Gamma$.}
 \end{figure}

\begin{figure}[htb]
 \centerline{\includegraphics [height=6.5cm,angle=0]{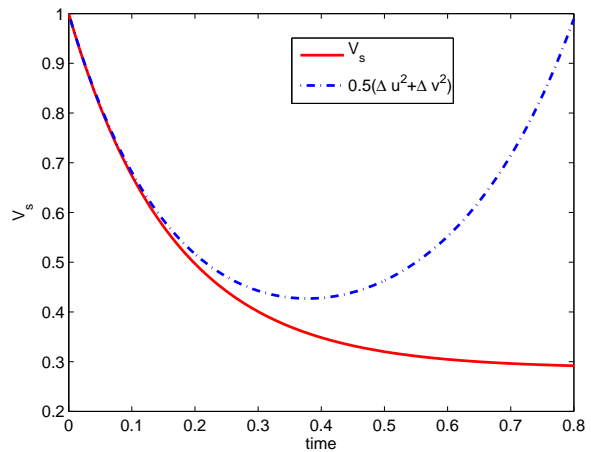}}
 \caption {\label{fig10} Plots of the time evolution of the smallest eigenvalue ($V_{s}$) and variance of the EPR operators ($\Delta u^{2}+\Delta v^{2}$) of the cavity radiation for $\kappa = 0.5$, $\theta=0.25$, $A=25$, $\Omega=10\Gamma$ and $\gamma=0.75\Gamma$.}
 \end{figure}

It is unequivocally shown in Fig. \ref{fig9} (please note that $A=25$ is used so that the dependence of the two can be distinctly observed) that the degree of entanglement that can be predicted by the two criteria exhibits very similar nature when $\Omega=0$. However, their nature turns out to be completely different when $\Omega\ne0$ in the longer time span as evinced in Fig. \ref{fig10} where the time parameter is limited to small interval so that the relation between the two can be readily noticeable. It is worth noting that $\Delta u^{2}+\Delta v^{2}$ rapidly increases with time which leads to dying of entanglement whereas $V_{s}$ continuously decreases with time.  What can be said at this juncture is that the two criteria predict similar nature of entanglement evolution for relatively smaller time scale, but they can also predict entirely different thing when there is an external driving radiation at longer time span. By the way, such a disparity between the two is not uncommon for $\eta=0$ which includes the present case \cite{jpb42215506}.

\subsection{Hillery-Zubairy criteria}

In earlier communication, on account of the relation of the criterion introduced by Hillery and Zubairy \cite{prl96050503} with the correlation of the photon number, the generated entanglement has been tipped to be quantifiable via simultaneous two-photon count measurement \cite{pra79063815}. In search of more appropriate and easy means of detecting entanglement, the degree of entanglement that can be predicted by the logarithmic negativity and HZ criteria has been compared earlier for nondegenerate three-level laser when $\Omega=0$, $\theta=0$ and $\gamma=\Gamma$. It was shown that both criteria predict the manifestation of entanglement at steady state, even though the degree of nonclassicality cannot be compared since HZ-criteria has no upper limit. In this contribution, upon changing the above parameters, the comparative evolution of the entanglement is investigated.

According to  HZ criteria, one can say that there is entanglement if
\begin{align}\label{p43}|\langle\hat{a}(t)\hat{b}(t)\rangle|>\sqrt{\langle\hat{N}_{a}(t)\rangle\langle\hat{N}_{b}(t)\rangle},\end{align} where $\langle\hat{N}_{a}(t)\rangle=\langle\hat{a}^{\dagger}(t)\hat{a}(t)\rangle$ and $\langle\hat{N}_{b}(t)\rangle=\langle\hat{b}^{\dagger}(t)\hat{b}(t)\rangle$ are the pertinent photon numbers corresponding to the involved modes, and $\langle\hat{a}(t)\hat{b}(t)\rangle$ is the intermodal correlation. Since the operators are already put in the normal order, Eq. \eqref{p43} can be expressed in terms of $c$-number variables associated with the normal ordering as
\begin{align}\label{p44}\langle\alpha(t)\beta(t)\rangle^{2}>\langle\alpha^{*}(t)\alpha(t)\rangle\langle\beta^{*}(t)\beta(t)\rangle.\end{align}

For nondegenerate three-level cascade laser, based on the assumption that there is no interaction among injected atoms, the photon number correlation is found to have the form \cite{pra74043816}
\begin{align}\label{p45}g(n_{a},n_{b})=1+{\langle\alpha(t)\beta(t)\rangle^{2}\over\langle\alpha^{*}(t)\alpha(t)\rangle\langle\beta^{*}(t)\beta(t)\rangle}.\end{align}
Hence upon comparing Eqs. \eqref{p44} and \eqref{p45}, one can readily see that the HZ inseparability criterion can be restated as
\begin{align}\label{p46}g(n_{a},n_{b})>2.\end{align}

It is obvious to see from  Eq. \eqref{p46} that $g(n_{a},n_{b})>2$ for all parameters under consideration. It is also clearly indicated that the evolution of the generated entanglement exhibits oscillatory nature. At this juncture, it is essential to note that for earlier time span $g(n_{a},n_{b})$ goes to infinity in connection with the mean photon number in one of the modes rapidly close to zero. In earlier report, similar behavior of  $g(n_{a},n_{b})$ has been explained in relation to the correlation involved in the criteria \cite{jpb42215506}. In light of this, it is possible to infer that except near the beginning of lasing operation, the two criteria unequivocally assert that the system under consideration generates entangled light. The overlap between these predictions may be utilized in detecting entanglement using the simultaneous photon count measurement in case the usual homodyne detection is found to be practically challenging.

\section{Conclusion}

Detailed analysis of the time evolution of the entanglement generated from a correlated emission laser is presented. Aiming at searching for a more reliable outcome, the entanglement criteria associated with the positivity of the transpose of the density matrix, the variance of the EPR type quadrature operators and photon number correlation are employed. Based on the assumption that various processes affect the preparation of the coherent superpositions and subsequent dynamics, the degree by which the phase fluctuates and the rate at which the coherence decays are taken as the probing mechanism. It is generally observed that in the absence of an external driving radiation ($\Omega=0$), when the phase can be locked ($\gamma<\Gamma$) and/or due to imperfect preparation when there is phase fluctuations ($\theta\ne0$), the generated entanglement  exhibits some sort of damping oscillatory nature. Upon varying the rate at which the coherent superposition decays and the deviation of the phase fluctuation, the degree of entanglement is found to be altered significantly particularly in the early stages of the lasing process.

However, for strong external driving radiation, the oscillatory nature of the quantum features and statistical properties of the radiation disappear. This is mainly because, the strong external driving radiation induces additional coherence while the  atoms traverse the cavity which subsequently compensates for the effect of dephasing and phase fluctuations. This explanation can be vividly evident if one  compares the effect of the two when $\Omega=0$ and $\Omega=10\Gamma$. Moreover, the oscillatory nature of the nonclassical features and photon statistics disappears for longer time span, since the steady state effects start to dominate. 

Critical scrutiny reveals that physically valid result can be obtained by carefull selection of these parameters. This can be related to the fact that when entirely viewed from physical perspective, the rate of dephasing and the phase fluctuation cannot be set arbitrarily since these physical processes are not entirely independent. This is mainly because, setting $\gamma/\Gamma$ small principally implies that the rate at which the coherent superposition decays is small. This, in other words, means that the phase between the upper and lower energy levels is locked by some degree. This does not arbitrarily go with the assumption that the phase fluctuates vigorously (large phase fluctuations). What possibly can be done in practical  utilization of the system is adjusting the two parameters in such away that one assumption does not preclude the effect of the other.

Comparison among the prediction of the nature of the evolution of the entanglement by various criteria more or less indicates the same thing. At this juncture, it would be appropriate citing the disparity between the prediction following from the criteria of logarithmic negativity and DGCZ for large $\Omega$ at longer time scale. By and large, this study evinces that the system under consideration can be a source of strongly entangled light for wide selection of parameters. Even though detailed analysis at steady state is still lacking, it is possible to infer that at longer time scale, externally driving the cavity significantly improves the degree of nonclassical features of the generated radiation.

\section*{Acknowledgments}

I thank the Max Planck Institute for Physics of the Complex Systems for allowing me to visit them and use their facility in carrying out this research and Dilla University for granting the leave of absence.


\begin{thebibliography}{1}
\bibitem{pra74043816} Tesfa S 2006 {\it{Phys. Rev. A}} {\bf{74}}   043816
\bibitem{jpb402373}  Tesfa S 2007 {\it{J. Phys. B: At. Mol. Opt. Phys.}} {\bf{40}}   2373 
\bibitem{pra72022305}  Han H T,  Zhu S Y and  Zubairy M S 2005 {\it{Phys. Rev. A}} {\bf{72}} 022305
\bibitem{jpb41145501} Tesfa S 2008 {\it{J. Phys. B: At. Mol. Opt. Phys.}} {\bf{41}} 145501
\bibitem{prl94023601}  Xiong H,  Scully M O and  Zubairy M S 2005 {\it{Phys. Rev. Lett.}} {\bf{94}} 023601
\bibitem{pra75033816}  Kiffner M,  Zubairy M S,  Evers J and  Keitel C H 2007 {\it{Phys. Rev. A}} {\bf{75}} 033816
\bibitem{pra79013831}  Qamar S,  Al-Amri M and  Zubairy M S 2009 {\it{Phys. Rev. A}} {\bf{79}} 013831
\bibitem{sinta}  Tesfa S 2010 {\it{Phys. Rev. A}} {\bf{82}} 053835 
\bibitem{sint}  Tesfa S 2010 arXiv:1011.3673
\bibitem{oc283781}  Qamar S,  Qamar S and  Zubairy M S 2010 {\it{Opt. Commun.}} {\bf{283}} 781
\bibitem{pra444688}  Majeed M and  Zubairy M S 1991 {\it{Phys. Rev. A}} {\bf{44}} 4688
\bibitem{pra79063815} Tesfa S 2009 {\it{Phys. Rev. A}} {\bf{79}} 063815
\bibitem{pra79033810}  Tesfa S 2009 Phys. Rev. A {\bf{79}} 033810
\bibitem{p149}  Susskind L and  Glogower J 1964 {\it{Physics}} {\bf{1}} 49;  Gerry C C 1987 {\it{Opt. Commun.}} {\bf{63}} 278;  Lynch R 1988 {\it{Opt. Commun.}} {\bf{67}} 67;  Lakshmi P A and  Swain S 1990 {\it{Phys. Rev. A}} {\bf{42}} 5632
\bibitem{pra77062308}  Qamar S,  Ghafoor F,  Hillery M and  Zubairy M S 2008 {\it{Phys. Rev. A}} {\bf{77}} 062308
\bibitem{jpb41055503} Tesfa S 2008 {\it{J. Phys. B: At. Mol. Opt. Phys.}} {\bf{41}} 055503 
\bibitem{pra75062305}  Qamar S,  Xiong H and  Zubairy M S 2007 {\it{Phys. Rev. A}} {\bf{75}} 062305
\bibitem{jpb41145504}  Lee S Y,  Qamar S,  Lee H W and  Zubairy M S 2008 {\it{J. Phys. B: At. Mol. Opt. Phys.}} {\bf{41}} 145504
\bibitem{prl771413}  Peres A 1996 {\it{Phys. Rev. Lett.}} {\bf{77}} 1413;   Mancini S,  Giovannetti V,  Vitali D and  Tombesi P 2002 {\it{Phys. Rev. Lett.}} {\bf{88}} 120401;  Raymer M G,  Funk A C,  Sanders B C and  de Guise H 2003 {\it{Phys. Rev. A}} {\bf{67}} 052104;  Doherty A C,  Parrarilo P A and  Spedalieri F M 2004 {\it{Phys. Rev. A}} {\bf{69}} 022308;  Hyllus P,  Guhne O,  Bruss D and  Lewenstein M 2005 {\it{Phys. Rev. A}} {\bf{72}} 012321;  Nha H and  Zubairy M S 2008 {\it{Phys. Rev. Lett.}} {\bf{101}} 130402
\bibitem{jpb42215506} Tesfa S 2009 {\it{J. Phys. B: At. Mol. Opt. Phys.}} {\bf{42}}   215506
\bibitem{pra65032314}  Vidal G and  Wener R F 2002 {\it{Phys. Rev. A}} {\bf{65}} 032314
\bibitem{prl842722}  Duan L M,  Giedke G,  Cirac J I and  Zoller P 2000 {\it{Phys. Rev. Lett.}} {\bf{84}}   2722
\bibitem{prl96050503}  Hillery M and  Zubairy M S 2006 {\it{Phys. Rev. Lett.}} {\bf{96}} 050503 
\bibitem{method} Barnett S M and  Badmore P M 1997 {\it{Methods in theoretical quantum optics}} (Oxford University Press, New York)
\bibitem{pla2231}  Horodecki M,  Horodcki P and  Horodecki R 1996 {\it{Phys. Lett. A}} {\bf{223}} 1;  Simon R 2000 {\it{Phys. Rev. Lett.}} {\bf{84}} 2726;  Giovannetti V,  Mancini S,  Vitali D and  Tombesi P 2003 {\it{Phys. Rev. A}} {\bf{67}} 022320;  Shchukin E and  Vogel W 2005 {\it{Phys. Rev. Lett.}} {\bf{95}} 230502;  Serafini A 2006 {\it{Phys. Rev. Lett.}} {\bf{96}} 110402;  Adesso G and  Illuminati F 2007 {\it{J. Phys. A: Math. Theor.}} {\bf{40}} 7821
\bibitem{job7577}  Laurat J,  Keller G,  Augusto O J,  Fabre C,  Coudreau T,  Serafim A,  Adesso G and  Illuminati F 2005 {\it{J. Opt. B: Quantum Semiclass. Opt.}}  {\bf{7}} 577 
\bibitem{pra70022318}  Adesso G,  Serafini A and  Illuminati F 2004 {\it{Phys. Rev. A}} {\bf{70}} 022318
\bibitem{prl93063601}  Fiurasek J and  Cerf N J 2004 {\it{Phys. Rev. Lett.}} {\bf{93}} 063601
\bibitem{jpb41215502}  Ma Y H,  Mu Q X,  Yang G H and  Zhou L 2008 {\it{J. Phys. B: At. Mol. Opt. Phys.}} {\bf{41}} 215502
\bibitem{pra79032334} Fujikawa K 2009 {\it{Phys. Rev. A}} {\bf{79}} 032334
\bibitem{pra77013815} Tesfa S 2008 {\it{Phys. Rev. A}} {\bf{77}} 013815 
\bibitem{pr47777}  Einstein A,  Podolsky B and  Rosen N 1935 {\it{Phys. Rev.}} {\bf{47}}
777 


\end{thebibliography}
\end{document}